\documentclass[10pt,letterpaper]{article}
\usepackage[top=0.85in,left=2.75in,footskip=0.75in]{geometry}

\usepackage{amsmath,amssymb}

\usepackage{changepage}

\usepackage[utf8x]{inputenc}

\usepackage{textcomp,marvosym}

\usepackage{cite}

\usepackage{nameref,hyperref}

\usepackage{microtype}
\DisableLigatures[f]{encoding = *, family = * }

\usepackage[table]{xcolor}

\usepackage{array}

\newcolumntype{+}{!{\vrule width 2pt}}

\newlength\savedwidth

\raggedright
\usepackage[aboveskip=1pt,labelfont=bf,labelsep=period,justification=raggedright,singlelinecheck=off]{caption}

\makeatletter
\renewcommand{\@biblabel}[1]{\quad#1.}
\makeatother

\usepackage{lastpage,fancyhdr,graphicx}
\usepackage{epstopdf}
\fancyheadoffset[L]{2.25in}
\fancyfootoffset[L]{2.25in}
\usepackage{graphicx}
\usepackage{subfigure}
\usepackage{epstopdf}
\usepackage{times}
\usepackage{bm}
\usepackage{color}

\graphicspath{{figs/}}

\begin{document}

\vspace*{0.2in}

\begin{flushleft}
{\Large
\textbf\newline{Spread of infectious disease and social awareness as parasitic contagions on clustered networks} 
}
\newline
\\
Laurent H\'ebert-Dufresne\textsuperscript{1,2,*},
Dina Mistry\textsuperscript{3},
Benjamin M.\ Althouse\textsuperscript{3,4,5}
\\
\bigskip
\textbf{1} Department of Computer Science, University of Vermont, Burlington, VT 05405, USA
\\
\textbf{2} Vermont Complex Systems Center, University of Vermont, Burlington, VT 05405, USA
\\
\textbf{3} Institute for Disease Modeling, Bellevue, WA, 98005, USA
\\
\textbf{4} University of Washington, Seattle, WA, 98105, USA
\\
\textbf{5} New Mexico State University, Las Cruces, NM, 88003, USA
\\
\bigskip

* laurent.hebert-dufresne@uvm.edu

\end{flushleft}

\section*{Abstract}
There is a rich history of models for the interaction of a biological contagion like influenza with the spread of related information such as an influenza vaccination campaign.
Recent work on the spread of interacting contagions on networks has highlighted that these interacting contagions can have counter-intuitive interplay with network structure.
Here we generalize one of these frameworks to tackle three important features of the spread of awareness and disease:
one, we model the dynamics on highly clustered, cliquish, networks to mimic the role of workplaces and households;
two, the awareness contagion affects the spread of the biological contagion by reducing its transmission rate where an aware or vaccinated individual is less likely to be infected;
and three, the biological contagion also affects the spread of the awareness contagion but by increasing its transmission rate where an infected individual is more receptive and more likely to share information related to the disease.
Under these conditions, we find that increasing network clustering,  which is known to hinder disease spread, can actually allow them to sustain larger epidemics of the disease in models with awareness.
This counter-intuitive result goes against the conventional wisdom suggesting that random networks are justifiable as they provide worst-case scenario forecasts.
To further investigate this result, we provide a closed-form criterion based on a two-step branching process (i.e., the numbers of expected tertiary infections) to identify different regions in parameter space where the net effect of clustering and co-infection varies.
Altogether, our results highlight once again the need to go beyond random networks in disease modeling and illustrate the type of analysis that is possible even in complex models of interacting contagions.

\section*{Author summary}
Epidemics of infectious diseases obviously interact with human behaviour, and mathematical models provide a principled way of studying these interactions. Unfortunately, the same scenario becomes less tractable when behaviour is affected by a social contagion, such as the \#FlattenTheCurve message currently spreading on social media in the hope of slowing the spread of COVID-19. Interactions across contagions are multi-dimensional: The messaging raises awareness and slows the spread of the disease around aware individuals by promoting hand washing and social distancing, and individuals who are experiencing the illness are also more likely to spread preventive messages. We model these interactions as a system of parasitic contagions where the awareness both benefit from and hinders the spread of the disease. We show how the dynamics of parasitic contagions differ from classic disease models, and most importantly, how social clustering or isolation can lead to worse outbreaks than the random mixing assumes by classic models. Our results illustrate the need to track messaging around public health crises and to include social awareness in our models and forecasts.



\section{Introduction} 

Models of contagion are used to study the transmission dynamics of a pathogen or information being transmitted through a structured population. 
Most of these are defined as compartmental models \cite{anderson1992infectious}, which mathematically distinguishes individuals based on their state; i.e., whether they are susceptible to a contagion or infectious with that contagion.
Using this approach, coupling different contagions to model their interactions is straightforward as we can then simply distinguish individuals based on all possible combinations of states for the different contagions.
Of particular interest is the coupling of an infectious disease with the spread of preventative information related to the disease \cite{funk2010modelling}. 
We typically expect this ``disease awareness'' to at least hinder, if not completely stop, the spread of the pathogen.
Since both the pathogen and the information spread through standard contagion mechanisms, their coupled dynamics is often referred to as either \textit{antagonistic} or \textit{dueling} contagions.
Here, we assume that the coupling between awareness and disease is twofold: Awareness reduces the transmission rate of a disease, but sees its own transmission rate increased by the presence of the disease.
Studies in health psychology suggest that individuals are more likely to adopt preventative behaviors related to a disease -- hand wawashing, self isolation, vaccination, treatment-seeking -- if people around them adopt that behavior and/or if they perceive themselves to be at high risk of infection \cite{Parker2013118,bruine2019reports}.
Indeed, data collected during and after the 2003 SARS coronavirus outbreak and the 2009 A (H1N1) influenza pandemic indicate the adoption of preventative behaviors such as increased hand washing, the use of face masks, and avoiding crowded public spaces in response to these outbreaks and the increased perception of infection risk \cite{Lau864SARS2003,Lau988SARS2003,Rubinb2651}.
We therefore refer to this coupled awareness-disease system as one of \textit{parasitic contagions}, since one contagion (awareness) both hurts and benefits from the other (the disease).

Research on dueling or parasitic contagions tend to assume well-mixed populations \cite{funk2009spread,funk2010modelling}, in line with classic epidemic models.
However, there are a number of generalized frameworks that account for the fact that information and disease both spread over a network structure \cite{funk2010interacting,marceau2011modeling,fu2017dueling}, and these frameworks have become more common in recent years with the rise in popularity of multiplex or multilayer network models \cite{granell2014competing,fan2016effect,scata2016impact,wang2016suppressing,zheng2018interplay}.
In the current work, we relax the well-mixed assumption to consider the community structure of most networks where connections are often grouped in dense environments such as households and workplaces.
To do so, we rely on two recent mathematical approaches: (i) a compartmental model coupled to a clique-based master equation formalism \cite{hebert2010propagation,hebert2015complex} that explicitly tracks the state of different groups; and (ii) a more straightforward two-step branching process analysis \cite{hebert2015complex} that approximates network clustering.  

In Section~\ref{sec:model} we present our general model of interacting contagion and provide the mean-field formalism to follow its evolution. We find that the impact of network clustering on the interaction of disease and awareness can be non-trivial to predict, and can even accelerate the spread of the disease. In Section~\ref{sec:complex}, we then develop a purely analytical tool based on a two-step branching process to better inform us of the net effect of clustering on the dynamics. The two approaches are leveraged together in a case study presented in Section~\ref{sec:results} where we show that there can be finite regions in parameter space where the disease can actually benefit from network clustering. Section~\ref{sec:conclusion} provides conclusions and outlines potential areas of future work. Altogether, our results raise potential questions about optimal coupling between the epidemiological parameters of a disease, the behavioral parameters of awareness, and the network structure of the population.

\section{Awareness and disease as parasitic infections\label{sec:model}}

\subsection{Network structures\label{sec:CS}}

\begin{figure}[t!!!]
\centering
\includegraphics[]{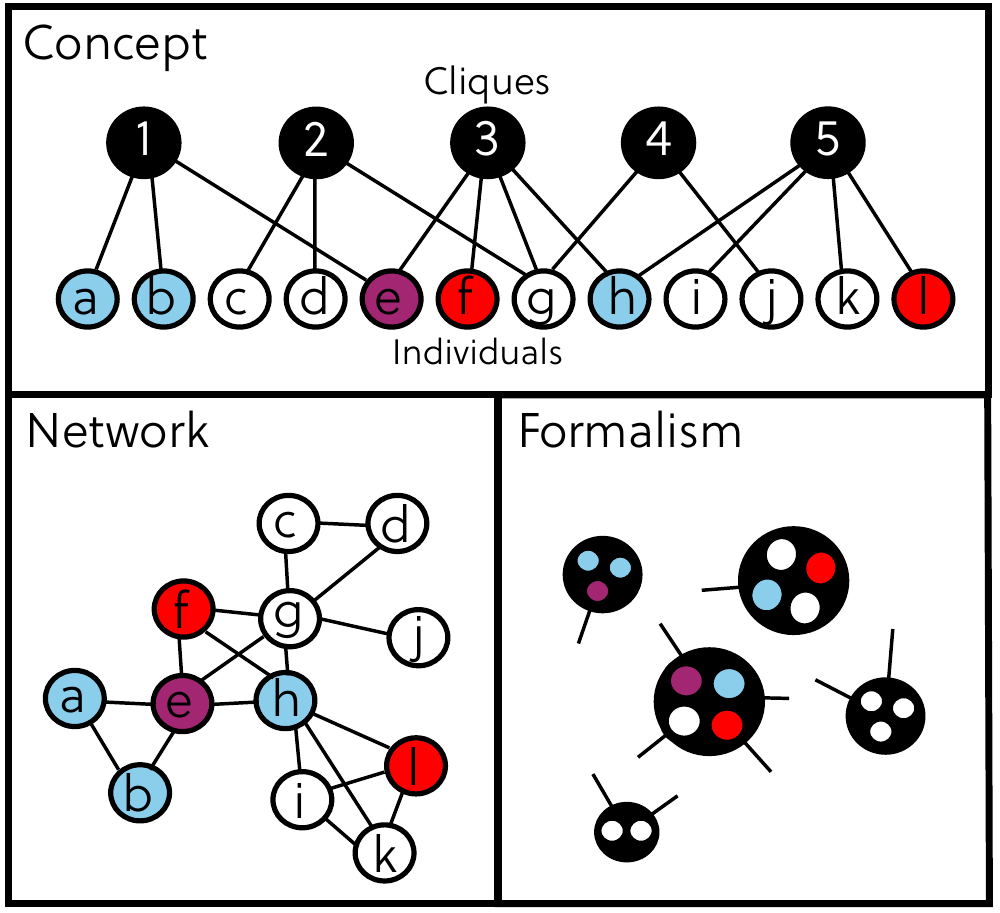}
\caption{Schematization of the particular topology and dynamics studied in this paper. An open circle represents a susceptible individual; a shaded one, a contagious individual (infected with the disease, awareness, or both); and a black circle represents a group (or clique). The topology is constructed by allowing individuals to belong to a given number of cliques where they can be linked to other participants (solid lines). Note that in the formalism, the cliques are distinguished by their exact population and state, while the precise connections between them remain unspecified. Modified from Ref.~\cite{hebert2010propagation}.}
\label{schema}
\end{figure}

To study parasitic contagions on clustered contact networks, we use a general definition of community structure where every network is decomposed in terms of groups \cite{newman03a}. The contact network between individuals can thus be interpreted as the projection of a bipartite networks where nodes are connected to social groups of different sizes. In this context, even random links are interpreted as groups of size two. The network topology of our model is illustrated in Fig.~\ref{schema}. In order to highlight the effects of community structure (CS) versus random network, the CS network will be compared with its equivalent random network (ERN): a network with exactly the same degree distribution, but with randomly connected nodes. Both topologies will be studied analytically and numerically.

Typical network datasets are often only available as a collection of pairwise edges rather than higher-order structure like groups. One then has to rely on some numerical methods such as community detection to infer group structure \cite{fortunato2016community}. Likewise, for theoretical models, one can simply rely on known distributions of groups per node (membership) and of nodes per group (group size) from previous studies on overlapping communities \cite{ahn2010link, xie2013overlapping}. We here use the simplest possible distributions in order to avoid confounding the impact of group structure from that of degree heterogeneity or degree correlations \cite{kiss2008comment}.

The dynamics of a single contagion on this community structure model was studied in \cite{hebert2010propagation}. Using a mean-field description, it was shown that the clustering of links in groups slowed down propagation as links are wasted on redundant connections instead of reaching new individuals. Expanding on this study, we more recently introduced a similar mean-field description for two synergistic disease \cite{hebert2015complex}, which is the model that we here generalize to interactions of other nature.


\subsection{Dynamical process}

To model the concurrent spread of an infectious disease and awareness of it on clustered networks, we will introduce a generalization of the model of interacting contagions used in Ref.~\cite{hebert2015complex}. We study the coevolution of two Susceptible-Infectious-Susceptible processes (SIS) such that, at any given time, the state of each individual is determined by their status regarding the two contagion processes. Without interaction with the other contagion, an individual with contagion $i$ would infect its susceptible neighbors at a rate $\beta _i$ and recover at a rate $\alpha_i$, but we will here introduce a parametrization scheme to modify these rates and model possible interactions as generally as possible. Note that our model is general and could be applied to any type of pairwise interaction between two SIS processes. However, as the notation will become quite involved we will ground our derivation by referring to the first contagion as the disease (with natural parameters $\beta_D$ and $\alpha_D$) and to the second contagion as awareness (with natural parameters $\beta_A$ and $\alpha_A$)

To keep track of both contagions simultaneously, we distinguish nodes by their state $\left[XY\right]_m$ where $m$ is their membership number, $X \in \{S_1,I_1\}$ corresponds to their state regarding the first contagion and $Y \in \{S_2,I_2\}$ their state regarding the second. Similarly, we will distinguish groups by their size $n$ and the states of the nodes they contain. I.e., $\left[ijk\right]_n$, where $i$ is the number of $\left[I_1S_2\right]$  nodes, $j$ is the number of $\left[S_1I_2\right]$ and $k$ is the number of $\left[I_1I_2\right]$; such that $n-i-j-k$ yields the number of $\left[S_1S_2\right]$. Keeping track of the number of nodes with both contagions is critical considering that we are interested in the effect of co-infection.

In the original model of Ref.~\cite{hebert2015complex}, co-infection had symmetric effect on both contagions, embodied in a single interaction parameter. For parasitic contagions, we want one contagion -- the awareness -- to benefit from being in the neighborhood of the other, the disease. Individuals might be more likely to listen to an awareness campaign if they are themselves sick or if the message comes from a sick individual. Likewise, they might be less likely to forget important information related to a disease if they are currently infected. We thus expect an increase in awareness transmission rate around infected individuals, and a decrease in loss of awareness for infected individuals. Second, we also want the disease to be hindered whenever nodes in a given neighborhood are aware of transmission risks or related treatment options. We might thus expect a decrease in disease transmission rate around aware individuals and/or an increase in disease recovery rate for aware individuals who might avoid contacts and seek treatment.

\begin{table}[t]

  \caption{\label{tab:hrn_networks}Description of all parameters in the model \vspace{0.15cm}}
  \begin{tabular}{l c }
    \hline
    \hline
    \multicolumn{1}{c}{Symbol$\quad$} & \multicolumn{1}{c}{Definition}  \\
        \hline
    $\lbrace g_m \rbrace$ & Distribution of groups per node (memberships)\\
    $\lbrace p_n \rbrace$ & Distribution of nodes per group (sizes)\\
        &\\  
    $\beta _D$ & Transmission rate of the disease ($S_1$ $\rightarrow$ $I_1$)\\
    $\beta _A$ & Transmission rate of  awareness ($S_2$ $\rightarrow$ $I_2$)\\
    $\alpha _D$ & Recovery rate of the disease ($I_1$ $\rightarrow$ $S_1$)\\
    $\alpha _A$ & Recovery rate of  awareness ($I_2$ $\rightarrow$ $S_2$)\\
        &\\  
    $\rho _{SS}^{II}$ & Factor of $\beta _D$ when the infector is aware\\
    $\rho _{SI}^{IS}$ & Factor of $\beta _D$ when the infectee is aware\\
    $\rho _{SI}^{II}$ & Factor of $\beta _D$ when infector \textit{and} infectee are aware\\
    $\gamma _{SS}^{II}$ & Factor of $\beta _A$ when the infector is sick\\
    $\gamma _{IS}^{SI}$ & Factor of $\beta _A$ when the infectee is sick\\
    $\gamma _{IS}^{II}$ & Factor of $\beta _A$ when infector \textit{and} infectee are aware\\
        &\\  
    $\tau _D$ & Factor of $\alpha _D$ when the infected is also aware\\
    $\tau _A$ & Factor of $\alpha _A$ when the infected is also sick\\
    \hline 
    \hline
  \end{tabular}
\end{table}

To track all these possible interactions, we therefore need to distinguish each possible infection by the state $\left[XY\right]$ of the infector and the state $\left[UV\right]$ of the infectee. The interaction of these states are embodied in a set of parameters, $\rho _{UV}^{XY}$, $\gamma _{UV}^{XY}$, $\tau _D$ and $\tau _A$. The first two give the factors affecting the transmission rates of the first and second contagion, respectively, when dealing with a $\left[XY\right]$ to $\left[UV\right]$ contact. For example, a $\left[I_1 I_2\right]$ individual will transmit the disease to a $\left[S_1 I_2\right]$ individual at a rate $\rho _{SI}^{II} \beta _D$. Of course, $\rho _{UV}^{XY} = 0$ whenever $U\equiv I$ and $\gamma _{UV}^{XY} = 0$ whenever $V\equiv I$ as these individuals are already infected with the corresponding contagion; similarly $\rho _{UV}^{XY} = 0$ whenever $X\equiv S$ and $\gamma _{UV}^{XY} = 0$ whenever $Y\equiv S$ as only infected individuals can transmit the contagion. We also consider that $\rho _{SS}^{IS} = \gamma _{SS}^{SI} = 1$ to preserve the natural transmission rate of each contagion, although we still use this term in the general equations. Finally, $\tau _D$ and $\tau _A$ give the factor by which the recovery rate of the disease or awareness are modified if the individual is also aware or sick, respectively.

\subsection{Mean-field description}

A mean-field description of the time evolution of our general model can be written in the spirit of previous formalisms. Leaving out all explicit mention of time dependencies as all variables and mean-fields vary in time, the population density within each node state evolves as
\begin{align}
\frac{d}{dt}\left[S_1S_2\right]_m & = \alpha_D\left[I_1S_2\right]_m + \alpha_A\left[S_1I_2\right]_m \nonumber \\
& - m\left(\beta_D B^{(D)}_{SS} + \beta_D B^{(A)}_{SS}\right)\left[S_1S_2\right]_m \label{eq:ss} \\
\frac{d}{dt}\left[I_1S_2\right]_m & = \tau _A \alpha_A\left[I_1I_2\right]_m - \alpha_D\left[I_1S_2\right]_m \nonumber \\
&  + m\left(\beta_A B^{(D)}_{SS}\left[S_1S_2\right]_m - \beta_D B^{(A)}_{IS}\left[I_1S_2\right]_m\right) \label{eq:is} \\
\frac{d}{dt}\left[S_1I_2\right]_m & = \tau_D \alpha_D\left[I_1I_2\right]_m - \alpha_A\left[S_1I_2\right]_m \nonumber \\
&  + m\left(\beta_A B^{(A)}_{SS}\left[S_1S_2\right]_m - \beta_D B^{(D)}_{SI}\left[S_1I_2\right]_m\right) \label{eq:si} \\
\frac{d}{dt}\left[I_1I_2\right]_m & = -(\tau_D \alpha_D+ \tau_A \alpha_A)\left[I_1I_2\right]_m \nonumber \\
&  + m\left(\beta_D B^{(D)}_{SI}\left[S_1I_2\right]_m + \beta_A B^{(A)}_{IS}\left[I_1S_2\right]_m\right)\label{eq:ii}
\end{align}
where $B^{(i)}_{UV}$ is a mean-field value of interactions representing the expected number of interactions with contagion $i$, per membership, for a node in state $\left[UV\right]$. Notice that in the equations, the first row of terms are the recovery events, and the second the infection events. The challenge in correctly writing the equations is thus solely to correctly identify to which state each event transfers some population density. Conservation of total population density (i.e. the sum over all state densities remains equal to one) is easily verified since the sum of Eqs.~(\ref{eq:ss}) to (\ref{eq:ii}) is zero.

Let us assume that we know the density $\left[ijk\right]_n$ of cliques that contain $n$ individuals with $i$ nodes contagious with the first contagion only, $j$ contagious with the second contagion only, and $k$ with both contagions. We could use this information to write the interaction mean-fields for the average level of interaction with contagious individuals within a given group:
\begin{align}
B^{(D)}_{SS} = & \frac{\sum _{\left[ijk\right]_n} \left(n\!-\!i\!-\!j\!-\!k\right)\left(i\rho_{SS}^{IS}+k\rho_{SS}^{II}\right)\left[ijk\right]_n}{\sum _{\left[ijk\right]_n}  \left(n\!-\!i\!-\!j\!-\!k\right)\left[ijk\right]_n} \\
B^{(D)}_{SI} = & \frac{\sum _{\left[ijk\right]_n} j\left(i\rho_{SI}^{IS}+k\rho_{SI}^{II}\right)\left[ijk\right]_n}{\sum _{\left[ijk\right]_n} j\left[ijk\right]_n} \\
B^{(A)}_{SS} = & \frac{\sum _{\left[ijk\right]_n} \left(n\!-\!i\!-\!j\!-\!k\right)\left(j\gamma _{SS}^{SI}+k\gamma _{SS}^{II}\right)\left[ijk\right]_n}{\sum _{\left[ijk\right]_n}  \left(n\!-\!i\!-\!j\!-\!k\right)\left[ijk\right]_n} \\
B^{(A)}_{IS} = & \frac{\sum _{\left[ijk\right]_n} i\left(j\gamma _{IS}^{SI}+k\gamma _{IS}^{II}\right)\left[ijk\right]_n}{\sum _{\left[ijk\right]_n} i\left[ijk\right]_n} \; .
\end{align}
These expressions can be understood with the following logic. For instance, in the case of $B^{(D)}_{SS}$, the susceptible individual is twice as likely to be part of a clique with twice as many susceptible nodes, which is what the $(n-i-j-k)$ factor takes into account. We then simply average the infection terms of each possible clique, i.e. $i\rho_{SS}^{IS}+k\rho_{SS}^{II}$, over this biased distribution of cliques.

With a variant of these mean-fields, we can now follow the evolution of group states by a general, but complicated, equation:

\begin{align}
\frac{d}{dt}\left[ijk\right]_n & = \left(i+1\right)\alpha_D\left[\left(i+1\right)jk\right]_n  + \left(j+1\right)\alpha_A\left[i\left(j+1\right)k\right]_n+ \nonumber \\
&\left(k+1\right) \times\bigg\lbrace \tau_D \alpha_D\left[i\left(j-1\right)\left(k+1\right)\right]_n+\tau _A \alpha_A\left[\left(i-1\right)j\left(k+1\right)\right]_n\bigg\rbrace \nonumber \\
 & - \left(i\alpha_D+j\alpha_A+k\tau_D \alpha_D+k\tau_A \alpha_A\right)\left[ijk\right]_n + \beta _D \left(n\! -\! i\! +\! 1\! -\! j\! -\! k\right) \times \nonumber \\
 &  \bigg\lbrace\left(i\! -\! 1\right)\rho_{SS}^{IS}+k\rho_{SS}^{II}+\tilde{B}^{(D)}_{SS}\bigg\rbrace \left[\left(i\! -\! 1\right)jk\right]_n \nonumber \\
 & - \beta_D\left(n\! -\! i\! -\! j\! -\! k\right)\bigg\lbrace i\rho_{SS}^{IS}+k\rho_{SS}^{II}+ \tilde{B}^{(D)}_{SS}\bigg\rbrace \left[ijk\right]_n  \nonumber \\
 &+ \beta _A \left(n\! -\! i\! -\! j\! +\! 1\! -\! k\right)\bigg\lbrace\left(j\! -\! 1\right)\gamma_{SS}^{SI}+k\gamma_{SS}^{II}+\tilde{B}^{(A)}_{SS}\bigg\rbrace \left[i\left(j\! -\! 1\right)k\right]_n \nonumber \\
 & - \beta_A\left(n\! -\! i\! -\! j\! -\! k\right)\bigg\lbrace j\gamma_{SS}^{SI}+k\gamma_{SS}^{II}+ \tilde{B}^{(A)}_{SS}\bigg\rbrace \left[ijk\right]_n  \nonumber \\
 & +  \beta_A \left(i\! + \!1\right)\bigg\lbrace j\gamma _{IS}^{SI} + \left(k-1\right)\gamma _{IS}^{II} + \tilde{B}^{(A)}_{IS}\bigg\rbrace \left[\left(i\! +\! 1\right) j \left(k\! -\! 1\right)\right]_n \nonumber \\
 & - \beta _A i \bigg\lbrace j\gamma _{IS}^{SI} + k\gamma _{IS}^{II} + \tilde{B}^{(A)}_{IS}\bigg\rbrace \left[ijk\right]_n \nonumber \\
 &  + \beta _D \left(j\! + \! 1\right)\bigg\lbrace i\rho_{SI}^{IS} + \left(k-1\right)\rho_{SI}^{II} + \tilde{B}^{(D)}_{SI}\bigg\rbrace \left[i \left(j\! +\! 1\right) \left(k\! -\! 1\right)\right]_n \nonumber \\
 & - \beta _D j \bigg\lbrace i\rho_{SI}^{IS} + k\rho_{SI}^{II} + \tilde{B}^{(D)}_{SI}\bigg\rbrace \left[ijk\right]_n
 \label{cliqueequation}
\end{align}
which is defined over all non-negative integers $n\geq 2$ and $i+j+k\leq n$. Eq. (\ref{cliqueequation}) is coupled to the previous system of ODEs through the mean-field values of \textit{excess} interactions $\tilde{B}^{(x)}_{UV}$, representing interactions with \textit{outside} groups, given by
\begin{equation}
\tilde{B}^{(x)}_{UV} = \left(\frac{\sum _m m(m-1)\left[UV\right]_m}{\sum _m m\left[UV\right]_m}\right) B^{(x)}_{UV} \; .
\end{equation}

The first four terms of Eq.~(\ref{cliqueequation}) are those corresponding to recoveries; positive for those corresponding to cliques relaxing into $\left[ijk\right]_n$ and negative for those where $\left[ijk\right]_n$ relaxes into a less infected state. The other terms represent each possible infection event. Notice that creating a $k$ individual implies either removing a $i$ or $j$, through their infection with contagion 2 or 1 respectively; just as recoveries can create $i$ or $j$ individuals when a $k$ individual recovers from contagion 2 or 1.

\begin{figure}[t!!!!!!!!!!!!!!]
\includegraphics[]{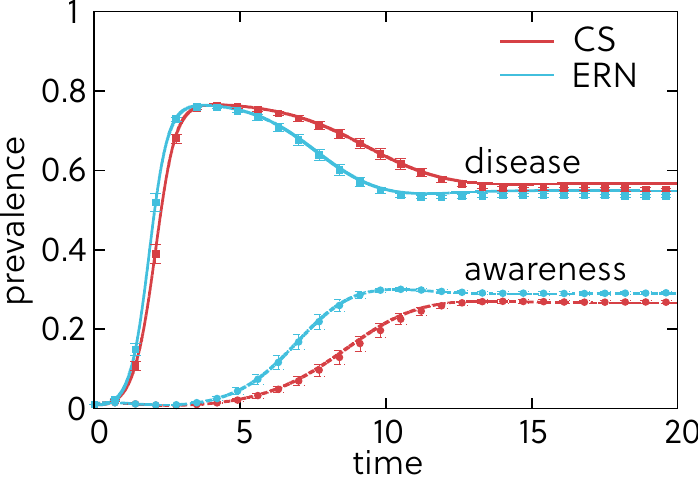}
\caption{\label{fig:validation} Parasitic interaction in a population where nodes all belong to two groups of 10 nodes. Markers represent average results of Monte Carlo simulations with error bars representing the standard deviation of over 100 runs on networks of 50 000 nodes. Solid curves are obtained by integrating the mean-field formalism. Results on the CS are shown in shade and those on its ERN are shown in black. The dynamics follow $\beta _D = 0.02$, $\beta_A = 0.25$, $\alpha_D = \alpha_A = \tau_D = \tau_A = 1.0$, $\rho_{UV}^{IS} = 1$, $\rho_{UV}^{II} = 10$ and $\gamma_{UV}^{XY} = 0.05$ except, $\rho_{SS}^{IS} = \gamma_{SS}^{SI} = 1.0$.}
\end{figure}

\subsection{Validation}

To validate the accuracy of our mean-field description, we run simulations on highly clustered networks where every node belongs to 2 cliques of size 10. We use this network for two reasons: First, to avoid degree-degree correlations, such that we know that the effect of clustering will be the main structural effect. Second, to feature a realistic local clustering coefficient, C, i.e. the ratio of triangles to pairs of links around a given node, which is here $C = 0.47$.
In Fig.~\ref{fig:validation}, we show prevalence for the disease and awareness over time on a clustered (CS) and an exponential random graph (ERN), using both Monte Carlo simulations and our ODE system. The accuracy of the mean-field approximations were expected given Refs.~\cite{hebert2010propagation,hebert2015complex}, and for the rest of the paper we therefore rely on the ODE system rather than slower Monte Carlo simulations. 

Most importantly, while we know that an awareness campaign or network clustering can both hinder the spread of a disease, it appears that network clustering can actually help a disease spread further when it is competing against a second contagion such as an awareness campaign. This result shows once more that the impacts of different dynamical or structural features can combine in non-trivial ways in models of contagion on networks.

\subsection{Two-step branching process\label{sec:complex}}

In Ref.~\cite{hebert2015complex}, we also introduced a simple criterion to determine whether a clustered structure would spread two synergistic contagions faster. This criterion can conceptually be thought of as a generalization of the basic reproductive number ($R_0$, the number of secondary infections from an average infectious individual in a completely susceptible population), which is often used to characterize the initial speed of epidemics, which in our case considers two infection steps in order to include clustering. Physically, it can be interpreted as a two-step branching process, as we count the number of tertiary infections caused by a co-infected individual (i.e. how many second neighbors will be infected). That being said, the analogy is imperfect: The ``branching process'' does not repeat itself since we do not distinguish which contagion(s) caused those tertiary infections. Yet, it proved to be a useful tool in Ref.~\cite{hebert2015complex} to identify the net effect of clustering across parameter space; i.e., to determine whether clustering speeds up or slows down propagation.

We start with a single node infected with both contagions, and denote the average excess degree of recently infected nodes as $z_1$ (i.e., we assume this node received the contagions from a single neighbor and $z_1$ describes the average number of \textit{other} neighbors this node is expected to have). For the first step of our criterion, we need to distinguish the probability of transmitting only the disease, only the awareness, or both. Since we ignore re-infection events, the latter scenario can occur in two ways: either by transmitting both while co-infected; or transmitting the first (or second) while co-infected before recovering from it and then transmitting the second (or first). Summing the two events yields the probability $T_{II}$ of a co-infected transmitting both contagions to a given first neighbor, i.e.,

\begin{align}
T_{II} = & \frac{\rho_{SS}^{II}\beta_D}{\rho_{SS}^{II}\beta_D+\gamma_{SS}^{II}\beta_A+\tau_D\alpha_D+\tau_A\alpha_A} \times \nonumber \\ 
& \left[\frac{\gamma_{IS}^{II}\beta_A}{\gamma_{IS}^{II}\beta_A+\tau_D\alpha_D+\tau_A\alpha_A}+\frac{\tau_D\alpha_D}{\gamma_{IS}^{II}\beta_A+\tau_D\alpha_D+\tau_A\alpha_A}\left(\frac{\gamma_{IS}^{SI}\beta_A}{\gamma_{IS}^{SI}\beta_A+\alpha_A}\right)\right] \nonumber \\
 & + \frac{\gamma_{SS}^{II}\beta_A}{\rho_{SS}^{II}\beta_D+\gamma_{SS}^{II}\beta_A+\tau_D\alpha_D+\tau_A\alpha_A} \times \nonumber \\ 
 &\left[\frac{\rho_{SI}^{II}\beta_D}{\rho_{SI}^{II}\beta_D+\tau_D\alpha_D+\tau_A\alpha_A}+\frac{\tau_A\alpha_A}{\rho_{SI}^{II}\beta_D+\tau_D\alpha_D+\tau_A\alpha_A}\left(\frac{\rho_{SI}^{IS}\beta_D}{\rho_{SI}^{IS}\beta_D+\alpha_D}\right)\right] \; .
\end{align}

Similarly, a co-infected node can transmit only the disease in two ways, either by infecting while co-infected then recovering before the transmitting awareness or by recovering from awareness before transmitting the disease. Again, summing these events gives the probability $T_{IS}$ of a co-infected transmitting only the disease. The same logic applies to the probability $T_{SI}$ of transmitting only the second contagion. We can thus write

\begin{align}
T_{IS} = & \frac{\rho_{SS}^{II}\beta_D}{\rho_{SS}^{II}\beta_D+\gamma_{SS}^{II}\beta_A+\tau_D\alpha_D+\tau_A\alpha_A} \times \nonumber \\
& \left[1-\frac{\gamma_{IS}^{II}\beta_A}{\gamma_{IS}^{II}\beta_A+\tau_D\alpha_D+\tau_A\alpha_A}-\frac{\tau_D\alpha_D}{\gamma_{IS}^{II}\beta_A+\tau_D\alpha_D+\tau_A\alpha_A}\left(\frac{\gamma_{IS}^{SI}\beta_A}{\gamma_{IS}^{SI}\beta_A+\alpha_A}\right)\right] \nonumber \\
 & + \frac{\tau_A\alpha_A}{\rho_{SS}^{II}\beta_D+\gamma_{SS}^{II}\beta_A+\tau_D\alpha_D+\tau_A\alpha_A}\left(\frac{\rho_{SS}^{IS}\beta_D}{\rho_{SS}^{IS}\beta_D+\alpha_D}\right) \; , \\
T_{SI} = & + \frac{\gamma_{SS}^{II}\beta_A}{\rho_{SS}^{II}\beta_D+\gamma_{SS}^{II}\beta_A+\tau_D\alpha_D+\tau_A\alpha_A} \times \nonumber \\
& \left[1-\frac{\rho_{SI}^{II}\beta_D}{\rho_{SI}^{II}\beta_D+\tau_D\alpha_D+\tau_A\alpha_A}-\frac{\tau_A\alpha_A}{\rho_{SI}^{II}\beta_D+\tau_D\alpha_D+\tau_A\alpha_A}\left(\frac{\rho_{SI}^{IS}\beta_D}{\rho_{SI}^{IS}\beta_D+\alpha_D}\right)\right] \nonumber \\
 & + \frac{\tau_D\alpha_D}{\rho_{SS}^{II}\beta_D+\gamma_{SS}^{II}\beta_A+\tau_D\alpha_D+\tau_A\alpha_A}\left(\frac{\gamma_{SS}^{SI}\beta_A}{\gamma_{SS}^{SI}\beta_A+\alpha_A}\right)
\end{align}

In its first neighborhood, we now know that a single co-infected individual will on average cause $z_1 T_{II}$ co-infections, $z_1 T_{IS}$ transmissions of the disease only, and $z_1 T_{SI}$ transmissions of the awareness only. We call those secondary infections. Our two-step branching process then looks at the number of tertiary infections, i.e. the number of transmission events of either contagions in the second neighborhood. 

In a clustered network, there is an overlap between the second neighborhood and the first, such that neighbors of the original co-infection can be infected during the second step of the process if they were not already. Let us consider one of the $z_1 T_{IS}$ first neighbors infected only with the disease and now trying to infect a susceptible node. We know that in its own first neighborhood, a number $(z_1-1) C \left( T_{II} + T_{IS} \right)$ of them are already infected with the same contagion ($z_1-1$ is an approximation, equal to its excess degree minus the targeted susceptible node). The fact that a fraction of its neighborhood is already infected by the root node is the negative impact of clustering on the dynamics. However, a number $(z_1-1)C \left( T_{II} + T_{SI} \right)$ are now also aware, such that they could transmit it to the node of interest and change its transmissibility. This is a potentially positive impact of clustering depending on the nature of the coupling between contagions (e.g. positive for the spread of awareness, negative for the disease itself).

Still considering the same first neighbor infected with the disease only, we need to know the rate at which it is co-infected by one of its $(z_1-1)C \left( T_{II} + T_{SI} \right)$ aware neighbors. Assume that we know the value of that rate, denoted $x_D$ for a co-infection to a diseased node, then the probability of co-infection before recovery would simply be $x_D/(x_D+\alpha_D)$. Since we can also write that probability as every node involved recovering before co-infection, we can require the following equality:

\begin{align}
\frac{x_D}{x_D+\alpha_D} = \left(\frac{\alpha_D+\alpha_A}{\gamma _{IS}^{SI}\beta_A + \alpha_D + \alpha_A}\right)^{(z_1-1)CT_{SI}}\left(\frac{\alpha_D+\tau_A \alpha_A}{\gamma _{IS}^{II}\beta_A + \alpha_D + \tau_A \alpha_A}\right)^{(z_1-1)CT_{II}} \; .
\end{align}

The same logic applies for co-infection involving a node that is aware but not sick. We can solve for the effective rates of co-infection through clustering, i.e. $x_D$ and $x_A$, and obtain

\begin{align}
x_D = & \alpha_D\bigg[\left(\frac{\alpha_D+\alpha_A}{\gamma _{IS}^{SI}\beta_A + \alpha_D + \alpha_A}\right)^{-(z_1-1)CT_{SI}}\left(\frac{\alpha_D+\tau_A \alpha_A}{\gamma _{IS}^{II}\beta_A + \alpha_D + \tau_A \alpha_A}\right)^{-(z_1-1)CT_{II}}\nonumber\\ 
& - \left(\frac{\alpha_D+\alpha_A}{\gamma _{IS}^{SI}\beta_A + \alpha_D + \alpha_A}\right)^{(z_1-1)CT_{SI}}\left(\frac{\alpha_D+\tau_A \alpha_A}{\gamma _{IS}^{II}\beta_A + \alpha_D + \tau_A \alpha_A}\right)^{(z_1-1)CT_{II}}  \bigg] \; , \\
x_A = & \alpha_A\bigg[\left(\frac{\alpha_D+\alpha_A}{\rho_{SI}^{IS}\beta_D + \alpha_D + \alpha_A}\right)^{-(z_1-1)CT_{IS}}\left(\frac{\alpha_A+\tau_D \alpha_D}{\rho_{SI}^{II}\beta_D + \tau_D \alpha_D + \alpha_A}\right)^{-(z_1-1)CT_{II}}\nonumber\\ 
&  - \left(\frac{\alpha_D+\alpha_A}{\rho_{SI}^{IS}\beta_D + \alpha_D + \alpha_A}\right)^{(z_1-1)CT_{IS}}\left(\frac{\alpha_A+\tau_D \alpha_D}{\rho_{SI}^{II}\beta_D + \tau_D \alpha_D + \alpha_A}\right)^{(z_1-1)CT_{II}} \bigg] \; .
\end{align}

With these effective rates, we can write the probabilities of a tertiary transmission of either disease or awareness, respectively $T_{IS}^{(D)}$, $T_{IS}^{(A)}$ if coming from a node initially infected only with the disease and $T_{SI}^{(D)}$ and $T_{SI}^{(A)}$ if coming from a node initially infected only with awareness. We write

\begin{align}
& T_{IS}^{(D)} = \left[1-C\left(T_{IS}+T_{II}\right)\right]\left[\frac{\rho_{SS}^{IS}\beta_D}{\rho_{SS}^{IS}\beta_D + \alpha_D + x_D}+\frac{x_D}{\rho_{SS}^{IS}\beta_D + \alpha_D + x_D}\left(T_{IS}+T_{II}\right)\right] \; \\
& T_{IS}^{(A)} = \left[1-C\left(T_{SI}+T_{II}\right)\right]\left[0 + \frac{x_D}{\rho_{SS}^{IS}\beta_D + \alpha_D + x_D}\left(T_{SI}+T_{II}\right)\right] \; \\
& T_{SI}^{(D)} = \left[1-C\left(T_{IS}+T_{II}\right)\right]\left[0 + \frac{x_A}{\gamma_{SS}^{SI}\beta_A + \alpha_A + x_A}\left(T_{IS}+T_{II}\right)\right] \; \\
& T_{SI}^{(A)} = \left[1-C\left(T_{SI}+T_{II}\right)\right]\left[\frac{\gamma_{SS}^{SI}\beta_A}{\gamma_{SS}^{SI}\beta_A + \alpha_A + x_A}+\frac{x_A}{\gamma_{SS}^{SI}\beta_A + \alpha_A + x_A}\left(T_{SI}+T_{II}\right)\right],
\end{align}
where the initial factor is the probability that a given neighbor is currently susceptible and where the two terms in brackets are respectively the probability of directly passing the correct contagion before co-infection, or of passing it after co-infection. More directly, we can write the probability of a tertiary infection of either disease or awareness from an individual who received both contagion from the original co-infected:

\begin{align}
T_{II}^{(D)} & = \left[1-C\left(T_{IS}+T_{II}\right)\right]\left(T_{IS}+T_{II}\right) \; , \\
T_{II}^{(A)} & = \left[1-C\left(T_{SI}+T_{II}\right)\right]\left(T_{SI}+T_{II}\right) \; .
\end{align}

From all of these, we write the number of tertiary infections of disease or awareness caused by an original co-infected individual as
\begin{align}
R_1^{(D)} = z_1^2 \left(T_{IS}T_{IS}' + T_{II}T_{II}'\right) \; , \\
R_1^{(A)} = z_1^2 \left(T_{SI}T_{SI}' + T_{II}T_{II}'\right) \; .
\end{align}

These two $R_1$ quantities are not generative numbers per se, as the process is not multiplicative for two reasons: (i) we count all tertiary infections, not only nodes in the same co-infected state as our original node and (ii) the next step in the process would imply facing clustering in both the second \textit{and} first neighborhood. Nevertheless, in Ref.~\cite{hebert2015complex}, it was shown that comparing $R_1$ values obtained using $C>0$ to that of an ERN with $C=0$ allows us to determine whether clustering slows down the dynamics ($R_1(C>0)<R_1(C=0)$) or speeds it up ($R_1(C>0)>R_1(C=0)$). 

\begin{figure*}
\includegraphics[width=\textwidth]{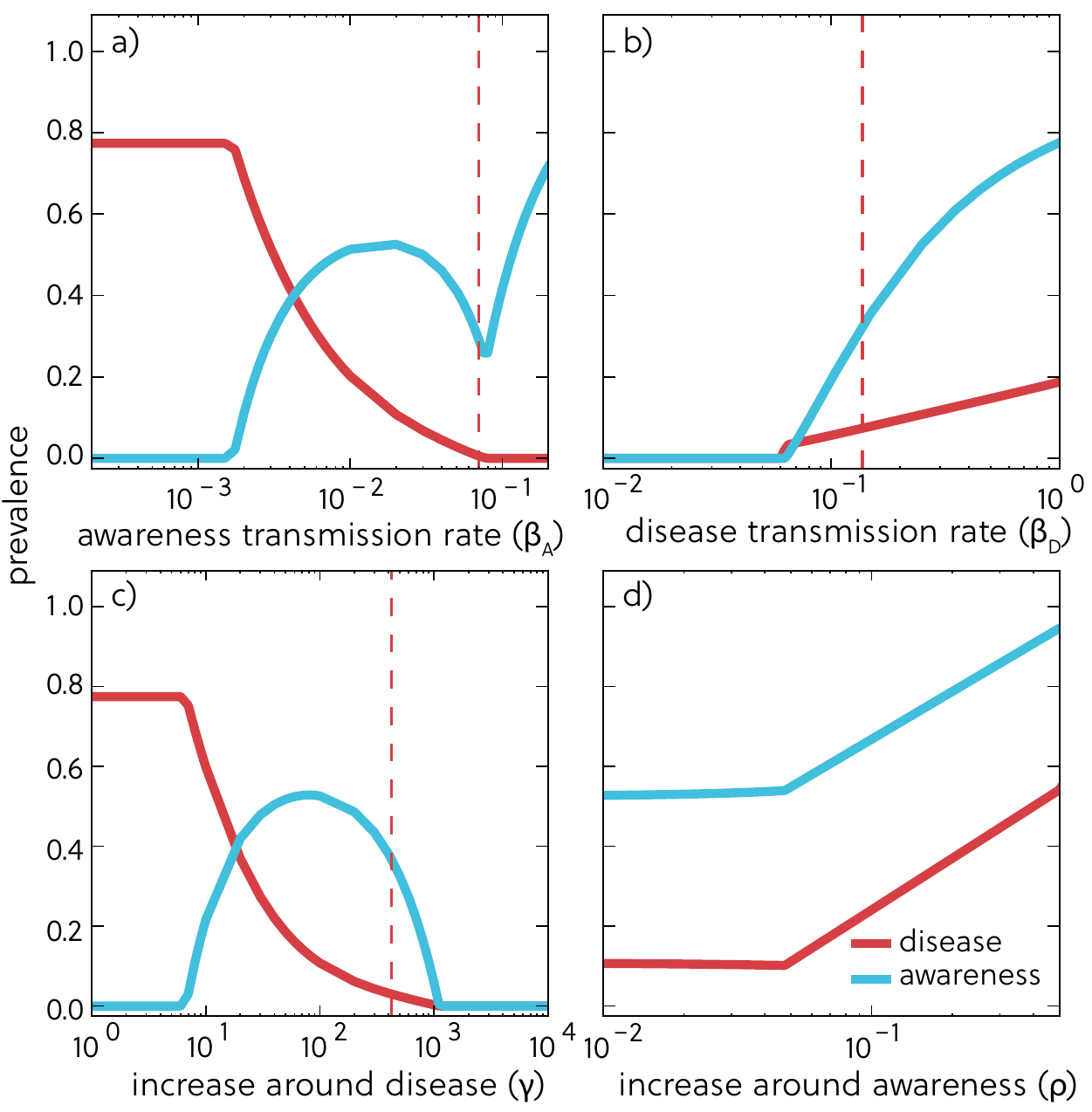}
\caption{\label{fig:bifurcation} Using the same clustered network structure and parametrization as in Fig.~\ref{fig:validation}, we now plot the final prevalence (i.e. final steady-state size) of disease and awareness across a range of parameters. a) We vary the transmission rate of awareness. b) We vary the transmission rate of the disease. c) We vary the increase in awareness transmission around sick individuals. d) We vary the decrease in disease transmission around aware individuals. The dotted vertical line marks the analytical epidemic threshold (if any) as approximated by $R_1^{(D)}=1$. All parameters are fixed to the following values unless we explicitly vary them: $\beta_D = 0.02$, $\beta_A = 0.25$, $\alpha_i = \tau_i = 1$, $\rho_{UV}^{IS} = 1$, $\rho_{UV}^{II} = \rho = 100$ and $\gamma_{UV}^{XY} = \gamma = 0.005$ except $\rho_{SS}^{IS} = \gamma_{SS}^{SI} = 1.0$.}
\end{figure*}

\begin{figure*}
\centering
\includegraphics[width=\textwidth]{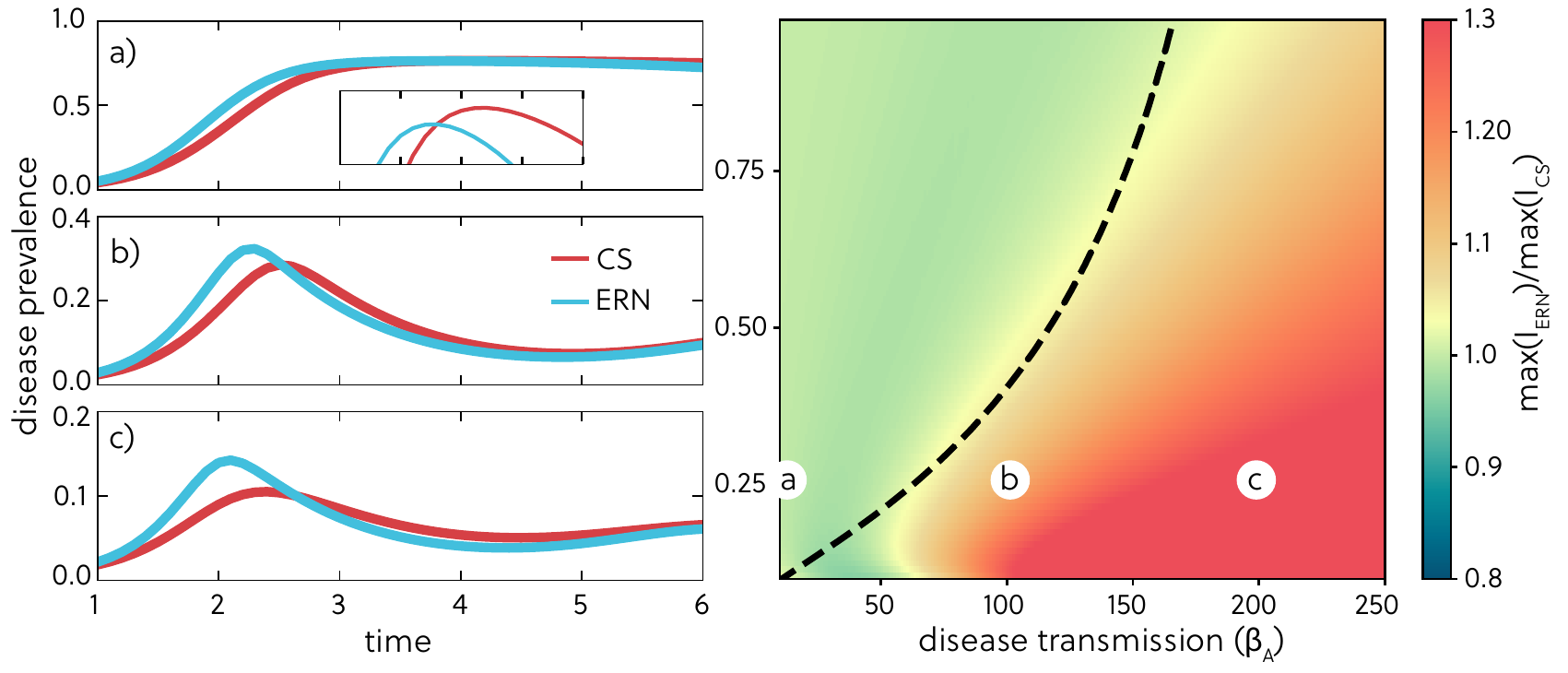}
\caption{Differences in peak prevalence values between network structure. We again use the same network structure and parametrization as in Fig.~\ref{fig:validation}, with a fixed $\beta_D = 0.02$, $\alpha_i = \tau_i = 1$, and with other interaction factors parametrized as $\rho_{UV}^{IS} = 1$, $\rho_{UV}^{II} = \rho$ and $\gamma_{UV}^{XY} = 1/(2\rho)$ except, $\rho_{SS}^{IS} = \gamma_{SS}^{SI} = 1.0$. Panels a-c) show the time series of disease prevalence for three sets of parameter values while panel d) study a large range of variations in $\beta_2$ and $\rho$. The inset of panel a) provides a closer look at the epidemic peak of both time series.
The color axis in panel d) shows the ratio of peak values obtained by integrating the mean-field system. The dotted line shows the crossover regime predicted by our analytical branching factor analysis; it defines the set of parameters for which $R_1^{(D)}/R_1^{(A)}$ is the same on both clustered and random networks. Finally, parameter values used in panels a-c) are shown with markers on panel d).}
\label{fig:peaks}
\end{figure*}

\section{Results\label{sec:results}}

In this section, we investigate the final epidemic sizes and peak prevalence values within the mean-field model, but also test the usefulness of the branching factor approach in identifying the epidemic threshold of the model as well as the net impact of clustering. Indeed, we can use the previous analysis to (i) identify whether the disease can maintain an outbreak despite clustering and awareness (i.e. if $R_1^{(D)}>1$) and (ii) evaluate whether a clustered network structure will lead to larger epidemic peaks than an equivalent random network (i.e. if $R_1^{(D)}/R_1^{(A)}$ is larger with $C>0$).

\subsection{Final sizes}

We investigate the robustness of the endemic disease state on clustered networks in Fig.~\ref{fig:bifurcation}. The most interesting case occurs when increasing the transmission rate $\beta_A$ of awareness can cause disease eradication by depleting the pool of susceptible individuals. At very low $\beta_A$, the awareness contagion fails to spread and the disease is left unhindered. At intermediate values of $\beta_A$, awareness is able to spread mostly due to its interaction with the disease, meaning it will reach a fraction of those already reached by the disease and fail to invade the susceptible population. In this regime, we find a non-monotonous relationship between the prevalence and transmission rate of awareness because increasing $\beta_A$ increases the probability of awareness reaching sick neighbors, while also decreasing the global fraction of sick individuals. After a certain threshold in $\beta_A$, the prevalence of the disease falls to zero and awareness then spreads as a regular contagion.

As shown in Fig.~\ref{fig:bifurcation}, the epidemic thresholds predicted by $R_1^{(D)}=1$ are typically within a factor 2 of the true epidemic threshold. While this is a good approximation, we find that in all cases the branching factor analysis systematically underestimates the robustness of the outbreak. This is most likely due to the fact that the analysis is seeded with a co-infected individual, while awareness and disease are likely to drift apart, benefiting the disease.

\subsection{Peak values}

In Ref.~\cite{hebert2015complex}, $R_1$ was used to determine whether two synergistic diseases would spread faster on a clustered or random network. The idea being that while clustering typically slows down dynamics, there can be an acceleration associated with the synergistic interactions and the benefit of being together by clustering. Here, both clustering in network structure and the interaction with awareness slow down the spread of the disease. We therefore do not expect to find a regime of accelerated disease spread. However, it is possible that clustering slows down awareness more than it slows down the disease, in which case a slower dynamics might still lead to a higher epidemic peak.

In Fig.~\ref{fig:peaks}, we vary the transmission rate of the disease and its interaction with awareness while tracking whether $R_1^{(D)}/R_1^{(A)}$ is larger with $C>0$ (larger epidemic peak on a clustered structure) or with $C=0$ (larger peak on the equivalent random network). We find that there can indeed be two separate regimes and the branching factor analysis provides a good approximation of where this crossover can occur.

\section{Outlook\label{sec:conclusion}}

With disease transmission comes the possibility for awareness of the disease and of the risk factors associated with its transmission. Awareness of the disease may cause individuals to respond by reducing their own transmissibility or adopting preventative behaviors. Here, we explored the effects of awareness in a model that looks at both disease and awareness as co-contagions in a parasitic relationship: spread of the disease leads to transmission of awareness which in turn leads to decreased disease prevalence as a result of reduced disease transmission around aware individuals. Our results show that interacting co-contagion models lead to different 
dynamics depending on the network structure on which they unfold. Characteristic measures 
such as the final outbreak size and the peak incidence exhibit regimes where they can be higher in networks exhibiting clustering than on equivalent but random network structures.

Altogether, our study highlights once again the need for disease models to go beyond random networks as social clustering can lead to either smaller or larger forecasts depending on the dynamics at play. We showed how interactions between contagions can combine with network structure in non-trivial ways and are therefore especially important to include in disease models. To this end, we have generalized the tools of Ref.~\cite{hebert2015complex} to account for more complicated interaction mechanisms.
In doing so, we end up with useful analytical tools, but their development becomes so involved and complicated as to be almost intractable. This raises the important problem of developing effective models for interacting contagions whose complexity does not grow exponentially with the number of contagions or with the number of interaction mechanisms. Indeed, not only do infectious diseases interact with social contagions such as vaccination and other preventative behaviors, but they also interact, often synergistically, with other biological infections \cite{halstead2003neutralization,shrestha2013identifying,nickbakhsh2019virus}. 
New tools are therefore needed to account for all of these interactions in a tractable and insightful analytical framework \cite{hebert2020macroscopic}.

Finally, with these theoretical advances also comes the need for improved data collection on the dynamics of awareness spreading and methods to measure how this materializes into effective preventative behaviors. With most of it now shared on online social media, information and messages regarding public health crises are increasingly important in shaping human behavior during epidemics. Unfortunately, data surrounding that messaging are not readily available to researchers and public health officials, even if we know it interacts in critical ways with our models and forecasts. The parallel development of theoretical frameworks and of data sharing protocols for social messaging related to public health crises will be invaluable going forward. Public awareness is an integral part of public health and advances to address its social media dimension should be integrated into existing public health surveillance systems.

\section*{Acknowledgments}
L.H.-D. acknowledges support from the National Institutes of Health 1P20 GM125498-01 Centers of Biomedical Research Excellence Award. D.M. and B.M.A. are supported by Bill and Melinda Gates through the Global Good Fund.

\pagebreak

\end{document}